\def\rfr#1{eq. (\ref{#1})}

\def\Rfr#1{Eq. (\ref{#1})}

\def\virg#1{``#1''}

\def\eqi{\begin{equation}}
\def\eqf{\end{equation}}
\def\eqia{\begin{eqnarray}}
\def\eqfa{\end{eqnarray}}
\def\rp#1#2{{#1\over#2}} \def\lb#1{\label{#1}}

\def\bds#1{\boldsymbol{#1}}

\documentclass{aastex}
\usepackage{url}
\usepackage{amsmath,amsthm,amscd,amssymb}
\usepackage{latexsym}
\usepackage{graphicx,epsfig}

\RequirePackage{color}

\newcommand{\emaila}{lorenzo.iorio@libero.it}

\begin{document}

\title{A-priori \virg{imprinting} of General Relativity itself on some tests of it?}
\shortauthors{L. Iorio}

\author{Lorenzo Iorio\altaffilmark{1} }
\affil{INFN-Sezione di Pisa. Permanent address for correspondence: Viale Unit\`{a} di Italia 68, 70125, Bari (BA), Italy.}

\email{\emaila}

\begin{abstract}
We investigate the effect of possible a-priori \virg{imprinting} effects of general relativity itself on satellite/spaceraft-based tests of it. We deal with some performed or proposed time-delay ranging experiments in the Sun's gravitational field. It turns out that the \virg{imprint} of general relativity on the Astronomical Unit and the solar gravitational constant $GM_{\odot}$, not solved for in the so far performed spacecraft-based time-delay tests, induces an a-priori bias of the order of $10^{-6}$ in typical solar system ranging experiments aimed to measuring the space curvature PPN parameter $\gamma$. It is too small by one order of magnitude to be of concern for the performed Cassini experiment, but it would affect future planned or proposed tests aiming  to reach a $10^{-7}-10^{-9}$ accuracy in determining $\gamma$.
\end{abstract}

\keywords{Experimental studies of gravity \*\ Ephemerides, almanacs, and calendars \*\ Lunar, planetary, and deep-space probes}
PACS: 04.80.-y, 95.10.Km, 95.55.Pe

\section{Introduction}
Several space-based tests of the Einstein's General Theory of Relativity (GTR) have been performed, or attempted, in the more or less recent past by following an \virg{opportunistic} approach, i.e. by suitably analyzing existing data sets of artificial satellites or interplanetary spacecrafts almost always built and launched for different original purposes (e.g. satellite geodesy, geodynamics, planetology, etc.). A cornerstone result was the Cassini radio science experiment \citep{Ies99} which lead to constraining the deviation of the PPN parameter $\gamma$ from its general relativistic value of unity at a  $10^{-5}$ level \citep{Ber03,And04} through the measurement of the general relativistic time delay affecting the electromagnetic waves linking the Earth and the Cassini spacecraft during its journey to Saturn when they were in superior conjunction, i.e. aligned and on opposite sides of the Sun.
 Many other high-precision space-based tests, aiming to reach accuracy levels as high as $10^{-7}-10^{-9}$ in determining $\gamma$, have been proposed \citep{Ni08,Ash09,Hob09,Tur09,Mil09}. Indeed, some theoretical scenarios predict deviations from GTR at such a level \citep{Nor1,Nor2,Dam1,Dam2}; for a a recent overview see, e.g., \citet{Tur}.

In this paper we wish to critically discuss certain subtle issues pertaining the consistency of such data analyses interpreted  as genuine tests of GTR.
In Section \ref{uno} we outline the problem in general terms. We will move to concrete examples in Section \ref{due} in which we will deal with the Cassini-like ranging experiments (Section \ref{dueuno}). Section \ref{tre} is devoted to the conclusions.
\section{The issue of the a-priori \virg{imprinting} of GTR itself in tests dedicated to it: general considerations}\lb{uno}
In general, in such tests huge data sets from man-made interplanetary probes/satellites are confronted to given dynamical models of their motion in which the relativistic effect to be tested was explicitly included, with one or more solve-for parameters $\{P\}$ accounting for it to be estimated in a least-square fashion along with many other ones $\{K\}$, not directly pertaining GTR.

Of crucial importance for interpreting such data analyses as genuine tests of GTR is to clarify how the numerical values of the models' parameters $\{F\}$  which have been kept fixed to certain reference values, i.e. which have not been solved-for, have been originally obtained.
From the point of view of testing GTR, it is not enough that the resulting post-fit residuals of a certain directly observable quantity are statistically compatible with zero at a good level; the standard data reduction procedure used for the original goals of the missions exploited for the GTR test considered may not be valid, in principle, for performing a truly unbiased, genuine check of GTR which is not a \virg{tautology}.

Indeed, if the primary task of a space-based mission is, for example, to reach a certain astronomical target with a given accuracy, the only thing that is important to this aim is that the dynamical models adopted to predict the probe's motion are accurate enough; this is usually quantitatively judged by inspecting the post-fit residuals of some directly measurable quantities like, e.g., ranges or range-rates. How the parameters $\{F\}$ entering the models have been obtained, i.e. their a-priori values, does not matter at all: the only important thing  is that the resulting fit of an existing set of  observations is good enough to minimize the observable's residuals.

Such  an approach may, in principle,  not be entirely adequate when the goal of the data analysis is testing a gravitational theory like GTR in an unambiguous, unbiased and self-consistent way. In this case, how the fixed parameters $\{F\}$  of the models have been obtained does, in fact, matter. Indeed, if one or more of them $\{I\}$ were previously obtained from different data of different bodies in such a way that they somehow retain a non-negligible a-priori \virg{imprint} of the same effect we are now interested in, their use may bias the current test just towards the desired outcome yielding, for example, a very high accuracy confirmation. In this cases, it would be more correct to use, if possible, values of such \virg{imprinted} parameters $\{I\}$ which have been obtained independently of the effect itself whose existence we are just testing in the present data analysis, even if the accuracy of such different values of the \virg{suspect} parameters $\{I\}$ was worse. Alternatively, if, for some reasons, such \virg{unbiased} values are not available, $\{I\}$ should be included, if possible, in the list of the solved-for parameters along with the one(s) $\{P\}$ accounting for the effect to be tested, and the resulting covariance matrix should be checked to inspect the correlations among them. The price to be paid may be an overall accuracy of the test not so high as that previously obtained\footnote{In principle,  certain confirmations may have their status changed.}, but we would have more epistemologically consistent, reliable and trustable tests.
\section{Application to some concrete cases}\lb{due}
\subsection{The Cassini radio-science test and other proposed high-accuracy space-based measurements of $\gamma$}\lb{dueuno}
To be more definite, let us look at the Cassini radio science test. In that case, the radiotechnical data of the spacecraft traveling to Saturn were contrasted with  a set of dynamical models by JPL of its motion and electromagnetic waves propagation in such a way that a correction $\Delta \gamma$ to the GTR-predicted value of the PPN parameter $\gamma$ was solved for, among other parameters, obtaining \citep{Ber03} \eqi\Delta\gamma\equiv|\gamma - 1|=(2.1\pm 2.3)\times 10^{-5};\eqf other authors got \citep{And04}
\eqi\Delta\gamma\equiv|\gamma - 1|=(-1.3\pm 5.2)\times 10^{-5}.\eqf

Now, a physical parameter which is, of course, crucial in such a test is the gravitational constant $GM_{\odot}$  of the Sun, which is the source of the relativistic time delay put on the test. It was not estimated \citep{Ber03,And04}, so that its numerical value was kept fixed to the standard reference figure of the JPL DE ephemerides used. It does, in principle, contain an a-priori \virg{imprinting} by GTR itself through the same effect itself that was just tested with Cassini, in particular by $\gamma$ itself. Indeed, the numerical value of $GM_{\odot}$ comes from the fixed value of the defining Gaussian constant\footnote{See on the WEB http://ssd.jpl.nasa.gov/?constants.}
\eqi k = 0.01720209895\ {\rm au^{3/2}\ d^{-1}},\eqf and from the value of the Astronomical Unit\footnote{Here we will use au for the symbol of the Astronomical Unit, like m for the meter, while AU will denote its numerical value in m.}, not estimated  in the Cassini tests,
\eqi {\rm AU}=1.49597870691\times 10^{11}\ (\pm 3)\ {\rm m}\eqf  through
\eqi GM_{\odot} = k^2\ {\rm AU}^3\ {\rm d}^{-2} = 1.32712440018\times 10^{20}\ (\pm 8\times 10^9)\ {\rm m^3}\ {\rm s^{-2}}.\lb{giemme}\eqf AU was, in fact, obtained just through a combination of radar ranging of Mercury, Venus, and Mars, laser ranging of the Moon (making use of light reflectors left on the lunar surface by Apollo astronauts), and timing of signals returned from spacecraft as they orbit or make close passes of objects in the solar system \citep{Sta04}; thus, it is affected in a non negligible way, given the level of accuracy of the techniques adopted,  by GTR itself and, in particular, by $\gamma$  which enters the PPN expressions for the time delay and bending of traveling electromagnetic waves.
Thus, there exists, in principle, the possibility that the high-accuracy results of the Cassini radio science tests may retain an a-priori \virg{imprint} of GTR itself through $GM_{\odot}$ (and the Astronomical Unit as well).

Let us put our hypothesis on the test by making some concrete calculations; for the sake of clarity, we will refer to the Cassini radio science tests, but the conclusions may be considered valid also for any of the many proposed $\gamma-$dedicated missions.

The GTR time delay experienced by electromagnetic waves propagating from point 1 to point 2 is
\eqi\Delta t = \rp{2R_g}{c}\ln\left(\rp{r_1 + r_2 + r_{12} + R_g}{r_1 + r_2 - r_{12} + R_g}\right),\lb{dela}\eqf
where $R_g=2GM_{\odot}/c^2$ is the Sun's Schwarzschild radius in which $G$ is the Newtonian gravitational constant, $M_{\odot}$ is the solar mass, $c$ is the speed of light in vacuum; $r_1$ is the heliocentric coordinate distance to point 1, $r_2$ is the heliocentric coordinate distance to point 2, and $r_{12}$ is the distance between the points 1 and 2. \Rfr{dela} is the expression actually used in the JPL'S Orbit Determination Program (ODP) used to analyze interplanetary ranging with planets and probes. In order to quantitatively evaluate the level of \virg{imprinting} by GTR itself in the used value of the Astronomical Unit, let us assume $r_1$ equal to the Earth-Sun distance and let us vary $r_2$ within 0.38 au and 1.5 au to account for the ranging to inner planets; the maximum effect occurs at the superior conjunction, i.e. when\footnote{Here $\bds n$ $\equiv \bds{r}/r $.}
${\bds{ n}}_1\approx -{\bds{n}}_2$, and $r_{12}\approx r_1 + r_2$. It turns out that $\Delta t_{\rm ranging}\approx 4\times 10^{-4}$ s, which is certainly not negligible with respect to the accuracy of the order of $10^{-8}$ s with which the light-time for\footnote{The value in km of the Astronomical Unit is obtained by measuring at a given epoch the distance between the Earth and a target body (a planet or a probe orbiting it)  by multiplying $c$ times the round trip travel time $\tau$ of electromagnetic waves sent from the Earth and reflected back by the target body, and confronting it with the distance, expressed in AU, between the Earth and the target body at the same epoch as predicted by some accurate dynamical ephemeris \citep{Sta04}.} 1 au $\tau_{\rm A}$ is actually measured (http://ssd.jpl.nasa.gov/?constants). As a consequence, the quantitative impact of the interplanetary ranging in the inner solar system to the determination of the Astronomical Unit is of the order
\eqi d{\rm AU}=c\Delta t_{\rm ranging} = 1.14291\times 10^5\ {\rm m},\lb{diau}\eqf not negligible with respect to the meter-level accuracy in measuring the Astronomical Unit; thus, $d$AU$/$AU$=8\times 10^{-7}$. Differentiating \rfr{giemme} with respect to au and \rfr{diau} yield
\eqi \rp{dGM_{\odot}}{GM_{\odot}}=2\times 10^{-6}.\lb{digiemme}\eqf Thus, we conclude that the technique adopted to determine the numerical values of the Astronomical Unit and of the Sun's $GM$ induced an a-priori \virg{imprint} of GTR on them of $8\times 10^{-7}$ and $2\times 10^{-6}$, respectively.

Let us apply this result to a typical radio science experiment in the solar system with $r_1$ fixed to the Earth-Sun distance. By writing $r_{1/2}= x_{1/2}$ au, with $x_{1/2}$ expressing distances in Astronomical Units, differentiation of \rfr{dela} with respect to au and $GM_{\odot}$, and \rfr{diau}-\rfr{digiemme} yield
an \virg{imprinting} effect of the order of\eqi \left.\rp{\delta(\Delta t)}{\Delta t}\right|_{\rm GTR} = 2\times 10^{-6}\eqf for $r_2$ up to tens\footnote{The Cassini test was performed with $r_2=7.43$ au \citep{Ber03}.} AU; it turns out that the largest contribution comes from $dGM_{\odot}$. It is too small by one order of magnitude with respect to the performed Cassini radio science tests, but it should be taken into account in the future, more accurate experiments whose expected accuracy is of the order of $10^{-7}-10^{-9}$, in the sense that the a-priori bias of GTR in the future determinations of deviations of $\gamma$ from unity will be as large as, or even larger than the effects one will to test, unless either $GM_{\odot}$ will be estimated as well along with $\gamma$ itself or a value obtained independently of it will be adopted.

About the first point, we mention that $\gamma$ and $\beta$ were never been estimated
in those solutions in which, among other things, also AU was fitted; instead, they were kept fixed \citep{Kra,Pit05,Fie08,StaPit,Fie09}.
Moreover, \citet{Fie09} used a modified version of their latest planetary ephemerides, named INPOP08d, in which they kept fixed AU and simultaneously fitted $GM_{\odot}$ to the observations along with other parameters, but they did not simultaneously estimate the PPN parameters as well which were kept fixed.
Instead, for the purposes outlined in this paper it would be important to simultaneously fit $\gamma$ and AU (or $GM_{\odot}$) and  inspect the covariance matrix for the correlation among AU (or $GM_{\odot}$) and $\gamma$.

Concerning the last point, a possible choice may consist, in principle, of using a figure for $GM_{\odot}$ obtained from measurements of the solar gravitational redshift
\eqi Z\equiv \rp{\nu_{\rm rec} - \nu_{\rm em}}{\nu_{\rm rec}}\approx \rp{GM_{\odot}}{R_{\odot}c^2}\eqf which, at first order, is independent of GTR\footnote{Indeed, it depends on the coefficient $g_{00}$ of the spacetime metric tensor; only the PPN parameter $\beta$ enters $g_{00}$ with the term of order $\mathcal{O}(c^{-4})$, at present undetectable. } itself. Latest measurements of the IR oxygen triplet 7772-7775, extrapolated to the Sun's limb \citep{Lop91}, yield an accuracy of the order of a few percent level. Projects to improve it with future missions have been proposed \citep{Cac06,Ber09}; for example, the use of the Magneto-Optical Filter technique, developed by \citet{Cac06}, would allow to reach a relative accuracy of $10^{-6}$. Anyway, it must be noted that extracting $GM_{\odot}$ from the measured gravitational red-shift also requires the knowledge of the Sun's radius $R_{\odot}$, which is uncertain at a $10^{-4}$ level.  Indeed, the  commonly accepted value for the solar radius was for a long time \citep{ALL73}
\eqi R_{\odot}=695.99\ {\rm Mm}\eqf
 $(1\ {\rm Mm}=10^6\ {\rm m})$, although it was not clear how such a figure was obtained; moreover, no error bar was released. Subsequent observations of solar $f-$mode frequencies induced \citet{Sch} and \citet{Ant}, who used data from the  Michelson Doppler Imager (MDI) mounted on the Solar and Heliospheric Observatory
 (SOHO) satellite and from the Global Oscillation Network Group (GONG) network, to conclude that the actual solar radius was $0.3-0.2$ Mm smaller: indeed, the estimate by \citet{Sch} is
 \eqi R_{\odot}=695.68\pm 0.02 \ {\rm Mm}.\eqf
 In \citet{BCD98} the value
 \eqi R_{\odot}=695.509\pm 0.026\ {\rm Mm}\eqf
 is reported from the Earth-based High Altitude Observatory's Solar Diameter Monitor campaign; it differs by about 0.5 Mm from the standard value by \citet{ALL73}, being inconsistent with it at much more than $3-\sigma$ level. In \citet{TG01} we find, from  SOHO/MDI frequency data for $p-$mode frequencies,
 \eqi R_{\odot}=695.69\pm 0.14\ {\rm Mm}.\eqf
 Note that the values by \citet{BCD98} and \citet{TG01} are mutually inconsistent, although at a $1-\sigma$ level only: the estimates by \citet{Sch} and \citet{BCD98} are not consistent at $3-\sigma$ level. The $f-$ mode and $p-$ mode measurements \citep{Sch, TG01} are, instead, consistent each other.  By the way, it can be noted that the discrepancies among the different best estimates are larger than the associated errors; thus, as a conservative evaluation of the accuracy in knowing the solar radius we will assume
 such differences.

\section{Conclusions}\lb{tre}
We have investigated the impact of possible a-priori \virg{imprinting} effects of GTR itself on satellite/spacecraft data analyses specifically designed  to test some general relativistic predictions. In particular, we considered the time-delay experiments conducted or proposed in the Sun's field with ranging to interplanetary spacecraft.

Concerning the time-delay tests, the  numerical values for the Astronomical Unit and the solar gravitational constant currently adopted in the ephemerides used so far to analyze spacecrafts' data retain an a-priori \virg{imprint} of GTR itself of the order of $8\times 10^{-7}$  and $2\times 10^{-6}$, respectively. As a consequence, the bias in typical solar system radio-science experiments is of the order of $10^{-6}$, which is one order of magnitude smaller than the accuracy level reached in the performed Cassini experiment, but it would be of concern for future planned tests aiming to measure deviations of the PPN parameter $\gamma$ from its general relativistic value at $10^{-7}-10^{-9}$ level.

In order to have genuine, unambiguous and unbiased tests of GTR which are not \virg{tautologic}, it would be necessary to either estimate the suspect parameters as well along with those accounting for the relativistic effect of interest or use values for them obtained independently from GTR itself.


\end{document}